\documentclass[letter,traditabstract,letterpaper]{aa}
\usepackage{graphicx}
\usepackage{txfonts}

\def\arcmin{$^{\prime}$}

\def\h2{H$_2$}
\def\n2h{N$_2$H$^+$}

\def\13co{$^{13}$CO}
\def\H13CO+{H$^{13}$CO$^+$}
\def\HCO+{HCO$^+$}
\def\c18o{C$^{18}$O}
\def\12co{$^{12}$CO}

\def\c+{C$^+$}

\def\h2{H$_2$}

\begin{document}

%\slugcomment{Not to appear in Nonlearned J., 45.}
\titlerunning{C$^+$ Detection of Warm Dark Gas in Diffuse Clouds } 
\authorrunning{Langer  et al.}

 \author{   
      W. D. Langer \and
      T.\,Velusamy \and
    J. L.\,Pineda \and
    P. F. Goldsmith \and
            D. Li. \and
        H. W. Yorke }

\offprints{W.\,D.\,Langer \email{William.Langer@jpl.nasa.gov}}
\institute{Jet Propulsion Laboratory, California Institute of Technology, 4800 Oak Grove Drive, Pasadena, CA 91109-8099, USA}

\title{C$^+$ detection of warm dark gas in diffuse clouds \thanks{Herschel is an ESA space observatory with science instruments provided
by European-led Principal Investigator consortia and with important participation from NASA.}
%\\(Version 15_r from :  \today )
}

\date{Received / Accepted }

\abstract { 
We present the first results of the {\it Herschel} open time key program, $\bf{G}$alactic $\bf{O}$bservations of $\bf{T}$erahertz $\bf{C^+}$ ({\bf GOT C+}) survey of the [CII] $^2$P$_{3/2}$--$^2$P$_{1/2}$  fine-structure line at 1.9 THz (158 $\mu$m) using the HIFI instrument on {\it Herschel}. We detected 146 interstellar clouds along sixteen lines--of--sight towards the inner Galaxy. We also acquired HI and CO isotopologue data along each line--of--sight for analysis of the physical conditions in these clouds. Here we analyze 29 diffuse clouds (A$_{\rm V}$ $<$ 1.3 mag.\,) in this sample characterized by having [CII] and HI emission, but no detectable CO. We find that [CII] emission is generally stronger than expected for diffuse atomic clouds, and in a number of sources is much stronger than anticipated based on their HI column density.  We show that excess [CII] emission in these clouds is best explained by the presence of a significant diffuse warm H$_2$, dark gas, component.  This first [CII] 158 $\mu$m detection of warm dark gas demonstrates the value of this tracer for mapping this gas throughout the Milky Way and in galaxies.}

%Using radiative transfer and excitation analysis estimates of the HI and CII emission, %context:
%{blank} 
%Aims
%{blank.}
%Method 
%{blank}
%Results 
%{blank}

\keywords{ISM: atoms ---ISM: molecules --- ISM: structure}

\maketitle

\section{Introduction}
\label{sec:introduction}

Interstellar gas plays a crucial role in the life cycle of Galaxies providing the material for star formation and as a repository of gas ejected by stars as part of their evolution.  
The diffuse atomic gas has been mapped in HI 21 cm surveys, and dense molecular H$_2$ clouds have been mapped indirectly with molecular tracers, primarily $^{12}$CO  (c.f. Dame et al. 2001) and in more limited surveys in its isotopologues.  Missing, however, is a widespread spectral tracer of H$_2$ not located in regions with conditions appropriate to form CO, the so-called ``dark gas" (Grenier et al. 2005).  One of the best candidates for this tracer is C$^+$, which is widely distributed in the Galaxy as shown by COBE FIRAS all sky (Bennett et al. 1994) and BICE inner Galaxy (Nakagawa et al. 1998) observations of its 158 $\mu$m  $^2$P$_{3/2}$--$^2$P$_{1/2}$ fine structure line.  In addition, [CII] is a density- and temperature-sensitive probe of diffuse clouds and photon dominated regions (PDRs). While COBE revealed that this line is the brightest far-IR line in the Galaxy, low spectral ($\Delta$v $\approx$ 1000 km s$^{-1}$) and spatial (7\degr) resolution could not reveal individual cloud components.  BICE data, while slightly better in this respect  (175 km s$^{-1}$ and 15\arcmin), is not adequate to resolve clouds along the line--of--sight (LOS).  Heterodyne receivers are critical to provide the necessary velocity resolution to identify and study individual clouds along the LOS; however, prior to operation of the HIFI instrument (de Graauw et al. 2010) on the {\it Herschel} (Pilbratt et al. 2010) only a handful of high spectral resolution [CII] spectra were available from bright HII regions (c.f. Boreiko \& Betz 1991). 

Here we report the first results of a large-scale survey of [CII] 1.9 THz line emission in the Galaxy being conducted under the {\it Herschel} open time key program, $\bf{G}$alactic $\bf{O}$bservations of $\bf{T}$erahertz $\bf{C^+}$ ({\bf GOT C+}).  The full {\bf GOT C+} program will observe about 900 LOS in the Galaxy, divided into four subprograms (Langer et al\,(2010): 1) a Galactic disk uniform volume sampling in longitude (covering all 360\degr), at b = 0, $\pm$0.5\degr, $\pm$1\degr, and $\pm$2$\degr$; 2) strip maps in the inner Galaxy; 3) a sampling of Heiles high--latitude clouds studied in absorption; and, 4) the Galactic warp in the outer Galaxy.  To date we have obtained high spectral resolution spectra of [CII] towards 16 LOS towards the inner Galaxy near longitude 340$\degr$ (5 LOS) and 20$\degr$ (11 LOS) taken during the performance verification phase and priority science program. In this small sample ($<$ 2\% of the complete {\bf GOT C+} survey) we detected 146 clouds in [CII] emission.  These data cover four broad categories of clouds as defined by their HI, [CII], and CO signatures: a) diffuse atomic clouds with high HI to [CII] intensity ratio and no detectable CO; b) diffuse molecular clouds with relatively low HI to [CII] intensity ratio (due to relatively strong [CII]) and no $^{12}$CO emission (A$_{\rm V}$$<$ 1.5 mag.\,); c) transition molecular clouds and PDRs detected in HI, [CII], and $^{12}$CO, but not $^{13}$CO; and, d) dense molecular clouds  detected with HI, [CII], $^{12}$CO, $^{13}$CO, and sometimes C$^{18}$O.  

In this $\it{Letter}$ we  analyze 29 diffuse clouds (A$_{\rm V}$$<$1.3 mag.) characterized by the presence of [CII] and HI emission, but having no CO emission. We find that [CII] emission is generally stronger than expected for diffuse atomic clouds (T $\sim$ 100 - 200\,K, n $\sim$ 100 - 200 cm$^{-3}$), and in about one-third of the sources [CII] emission is much stronger than anticipated.  This [CII] emission indicates the presence of warm ``dark gas" (H$_2$ without CO) associated with HI. Velusamy et al. (2010) discuss the transition C$^+$--$^{12}$CO clouds and their ``dark gas" envelopes, and Pineda et al. (2010) discuss the dense molecular cloud PDRs. 

%The observations are discussed in
%Section~\ref{sec:Observations} The comparison between our
%Section~\ref{sec:Results and Analysis} and we discuss our results in
%Section~\ref{sec:discussion}. We summarize our results in
%Section~\ref{sec:conclusions}.

\section{Observations}
\label{sec:observations}

%\begin{figure}[t]
\begin{figure*}[t]
\centering
\includegraphics[width=0.90\textwidth,angle=0]{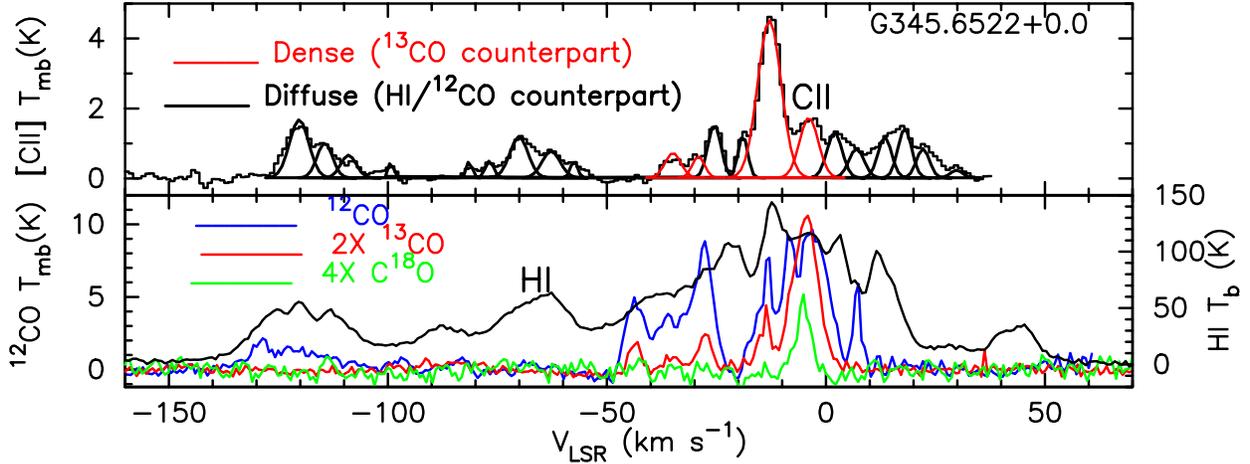}
\caption{[CII] spectra obtained with {\it Herschel} HIFI for the $\bf{GOT\,C+}$ program at $l$=345.65$\degr$ and b=0$\degr$, along with the CO data from the Mopra telescope and HI surveys. Many different types of interstellar clouds can be seen in this LOS; diffuse clouds with HI and C$^+$ emission, but without, or barely any, CO; strong [CII] emission towards dense molecular clouds, as indicated by the presence of $^{13}$CO and in some cases C$^{18}$O. The Gaussian decomposition for [CII] is also shown in the upper panel. }
%\label{fig:15088fg1}
\end{figure*} 
%\end{figure} 

The 16 LOS observations of [CII] at 1.9 THz were made with {\it Herschel} HIFI band 7b using Load Chop (HPOINT) with SkyRef offset by 2$\degr$ off the plane, for better cancelation of instrumental spectral baselines. The data were reduced with HIPE version 3 and a fringe-fitting tool to remove standing waves, and the h- and v-polarizations combined where available.  All [CII] data are corrected for the antenna's main beam efficiency of 0.63. We used the wide band spectrometer (WBS) with 0.22 km s$^{-1}$ at 1.9 THz.  The 16 LOS are: 1) b=0 at longitudes: 337.82$\degr$, 343.04$\degr$, 343.91$\degr$, 344.78$\degr$, 345.65$\degr$, 18.26$\degr$, 22.60$\degr$, 23.47$\degr$, 24.34$\degr$; and, 2) b = 0.5$\degr$ at 24.34$\degr$, b=-0.5$\degr$ at 18.26$\degr$, 23.47$\degr$, b=+1.0$\degr$ at 22.60$\degr$, 24.34$\degr$, and b=-1.0$\degr$ at 18.26$\degr$, 23.47$\degr$.
Further details are in Velusamy et al. (2010). We also observed the $J = 1 \to 0$ transitions of $^{12}$CO, $^{13}$CO, and C$^{18}$O toward each LOS  with the ATNF Mopra 22-m Telescope  (details in Pineda et al. 2010), with an angular resolution of 33$^{\prime\prime}$.  
We obtained HI data from public sources (McClure-Griffiths et al. 2005, Stil et al.\, 2006). 

A typical example of our data is shown in Figure 1 for l=345.6522$\degr$, b=0.0$\degr$, where we plot the [CII] antenna temperature (corrected for main beam efficiency) along with those of HI, and the CO isotopes (other examples are shown in Pineda et al.\,  2010 and Velusamy et al.\, 2010).  This LOS passes through the edge of the bar in the inner Galaxy.  The [CII] features are clearly blends of many cloud components.  It can be seen in Figure 1 that a wide variety of clouds are detected in [CII], for example: between V$_{lsr}$ = -20 and 10 km s$^{-1}$ there is strong [CII] emission along with CO and $^{13}$CO, whereas between -100 and -140 km s$^{-1}$ four narrow [CII] features appear which have weak $^{12}$CO associated with them, while between -50 and -90 km s$^{-1}$ there are five features with only [CII] and HI, and no detectable $^{12}$CO. 

\section{Results and analysis}
\label{sec:cii-comp-assoc}

We identified our clouds from Gaussian decompositions (see Figure 1) of [CII] and $^{13}$CO (where available) and use these to characterize their velocities, V$_{LSR}$, and linewidth $\Delta$v (see Velusamy et al.\, 2010 for details).  With this approach we detected [CII] in 146 components at the 3-$\sigma$ level or better.  We did not fit the HI spectral profiles, but  
obtained the HI intensities in each cloud for comparison with other lines, by integrating within the velocity width ($\Delta$v) centered at their V$_{LSR}$ defined by the [CII] (and $^{13}$CO) lines.  For each feature we derive the line parameters, T$_{peak}$(K), $\Delta$v(km s$^{-1}$), I = $\int$Tdv (units of K km s$^{-1}$)), as well as those for HI and CO (where detected).  
Here we focus on  29 out of 35 [CII] components that do not have any known molecular gas as traced by $^{12}$CO (the other six have marginal [CII] intensities, or other issues).  The linewidths (FWHM) for these 29 diffuse C$^+$ clouds, have a mean value of 3.4 km s$^{-1}$, and range from 1.4 to 5.5 km s$^{-1}$.  
In Figure 2 
we plot the integrated intensity I(CII) versus I(HI) for all 29 diffuse cloud features.  It can be seen that [CII] is not strongly correlated with I(HI).  This result is not what one would expect if all the C$^+$ is only in HI clouds. Furthermore, there is a large degree of scatter,  and, several sources have large [CII] emission at small N(HI), which we suggest is best explained by ``dark gas" -- H$_2$ without CO. 

The identification of ``dark gas" in our [CII] data can be understood by using simple models of the radiative transfer to relate the HI and [CII] intensities, and then consider the effects of adding an additional H$_2$ layer containing C$^+$. We calculate the column density of ionized carbon assuming the gas is only atomic, from the relation, N(C$^+$) = X(C$^+$) N(HI), where X(C$^+$) is the fractional abundance of C$^+$ in the gas with respect to HI. Observations in the local Interstellar Medium (ISM) lead to X(C$^+$) in the range 1.4  to 1.8 $\times$ 10$^{-4}$ (Sofia et al. 1997), and here we adopt a value of X(C$^+$)=1.5 $\times$ 10$^{-4}$.  The column density of HI can be estimated in the optically thin limit, where the brightness temperature is less than the kinetic temperature, as
$N({\rm HI}) =1.82{\times}10^{18} I({\rm HI}) \,{\rm (cm^{-2})}$. We can also directly calculate the N(C$^+$) column density from I(CII) if we know the temperature and density of the gas.  If the gas arises only from the HI cloud the N(C$^+$) solutions should be consistent within a reasonable range of density and temperature.  We derived the column density for N(C$^+$) in the optically thin limit (the effects of opacity are discussed below) with no background radiation field (the CMB is negligible), 
\begin{equation}
\quad
N_u({\rm C^+}) = \frac{8 \pi k \nu^2}{hc^3A_{ul}} \int{T_{mb}(CII)d{\rm v}} \,\,{\rm (cm^{-2})}\,
\end{equation}
where T$_{mb}$ is the main beam antenna temperature, N$_u$ is the upper level column density.  The total N(C$^+$) is derived by solving the excitation of the $^2$P$_{3/2}$ state (in the optically thin limit), 
%due to collisions and radiative decay.  
\begin{equation}
N({\rm C^+}) = 2.9\times10^{15} (1+0.5(1+n_{\rm cr}/n)e^{({\Delta}E/kT)}){\rm I}({\rm CII})\,{\rm (cm^{-2})}
\end{equation}
The critical density (for HI or H$_2$) is defined as, $n_{\rm cr} = A_{ul}/\left <\sigma {\rm v}\right>_{ul}$\,cm$^{-3}$, where A$_{ul}$ = 2.4$\times$10$^{-6}$ s$^{-1}$ and $\sigma$ is the collisional de-excitation cross-section for HI or H$_2$, respectively. For HI collisions with C$^+$ we used the $\left<\sigma {\rm v}\right>_{ul}$ calculations of Barinovs et al. (2005) and derived a fit, 3.8$\times$10$^{-10}$ T$^{0.15}$ cm$^3$ s$^{-1}$,  good to 20\% over the temperature range 50 - 300K, yielding $n_{\rm cr}(HI) = 3.1\times10^3$ cm$^{-3}$ at 100K.  The critical density for H$_2$ + C$^+$ de-excitation is twice that of HI (Flower 1988, 1990). 

In Figure 2 we plot the predicted I(CII) versus I(HI) as a function of temperature and density, assuming that N(C$^+$) is given by, N(C$^+$) = X(C$^+$)N(HI).  The solid lines show results for typical diffuse atomic cloud conditions,  T = 100 to 300K and n = 100 to 300 cm$^{-3}$.  The lowest curve is for (T,n) = (100K, 100 cm$^{-3}$), a likely lower bound for these diffuse atomic hydrogen clouds; all of our observed I(CII) fall above this line.  The result for somewhat denser diffuse clouds,  (T,n) = (100K, 300 cm$^{-3}$), is shown by the middle solid line in Figure 2; 
about 25$\%$ of our sources fall along this line. We also plot results for a very warm HI cloud at (T,n) = (300K, 300 cm$^{-3}$) (upper solid line) to give an idea of what physical conditions might be needed to explain a larger percentage of our [CII] clouds.  About half the observed [CII] clouds fall below this line.  However, this solution requires a rather hot diffuse atomic cloud, with a pressure, nT = 9$\times$10$^4$ that is much greater than the generally accepted value. At best it explains about half the [CII] sources as arising from atomic gas, but it is probably an unrealistically high combination of (n,T); and, the remaining sources require [CII] from additional cloud components. In summary, [CII] emission from diffuse HI gas alone does not explain that observed in much of our sample.

The analysis above assumes optically thin HI and [CII] emission.  We estimate the opacity correction for the [CII] lines by starting with the optically thin derived N(C$^+$) and using the RADEX code (van der Tak et al. 2007) to calculate $\tau$(CII) for a range of diffuse cloud densities and temperatures.  For I(CII) $\leq$ 5 K\,km\,s$^{-1}$, $\tau$  $\le$ 0.5 under most density and temperature conditions for a typical linewidth of 3 to 4 km s$^{-1}$, and the escape probability for [CII] photons is $\ge$ 80\%.  Thus, any corrections to the column density are $\le$ 20\%. The HI opacity is $\tau$(HI) =5.5$\times$10$^{-19}$ N(HI)$/$(T$_x$ $\Delta$v), or, using the optically thin estimate for N(HI) above, $\tau$(HI) =I(HI)$/$(T$_x$$ \Delta$v), where T$_x$ is the excitation temperature.  We do not know excitation temperature for HI, but it is reasonable to set it to 100 -- 150K, which yields $\tau$(HI) $\le$ 0.5, for I(HI) $\le$ 300 K\,km\,s$^{-1}$, and a maximum of 0.8 for the sources near I(HI) = 500 K\,km\,s$^{-1}$.  In summary, these corrections are small, and maintain approximately the same N(C$^+$)$/$N(HI) ratio as the optically thin case, which does not resolve the issue of excess [CII] emission within the context of purely atomic diffuse clouds.

Unless we adopt a very high density, n $ >$ 10$^3$ cm$^{-3}$ and$/$or high temperatures (T $>$ 300K), some other gas contribution is necessary to explain the [CII] emission.  The clouds with I(CII) $\ge$ 2.5 K\,km\,s$^{-1}$ and I(HII) $\le$ 300 K\,km\,s$^{-1}$ cannot even be explained by invoking such high HI densities and temperatures.  Furthermore, we would be faced with explaining how such dense clouds, not associated with HII regions, could have such high temperatures. We suggest that [CII] in these stronger C$^+$ emitting clouds, is tracing a significant component of warm molecular hydrogen gas, or warm ``dark gas".  Early cloud-chemical models  predicted such an H$_2$ gas layer with C$^+$ and little or no CO (see Langer 1977 and references therein), although the term ``dark gas" was not invoked until much later.  Since then, ever more accurate and sophisticated cloud models have been developed, but the basic structure of the HI to H$_2$ and C$^+$ to CO--CI profile remains the same (e.g. Tielens \& Hollenbach 1985, van Dishoeck \& Black 1988, Leeet al. 1996, Wolfire et al. 2010).

To estimate the amount of ``dark gas" needed to explain the relatively strong [CII] in diffuse clouds, we extended the excitation model of N(CII) versus I(CII) (equation 2) to a two-layer cloud, containing HI and H$_2$.  In the optically thin regime, which as shown above is a reasonable assumption for the diffuse sources, the total [CII] intensity is the sum of that in each layer,
 I(CII)${\rm _{total}}$= I(CII)${\rm_{HI}}$ + I(CII)${\rm_{H_{2}}}$.
%\begin{equation}
% I(CII)_{total} = I(CII)_{HI} + I(CII)_{H_{2}}.  
%\end{equation}
 We can solve for I(CI)$\rm_{HI}$ as follows: use 
 %equation (1) 
 $N({\rm HI}) =1.82{\times}10^{18}{\rm I(HI)} \,{\rm (cm^{-2})}$
 to solve for N(HI) from I(HI), then N(C$^+$) = X(C$^+$) N(HI); and, finally I(CII)${\rm_{HI}}$ can be determined from equation (2) assuming a temperature and density, (T$_{kin}$, n(HI)) in the HI layer. Since,
 I(CII)${\rm_{H_2}}$ =  I(CII)${\rm_{total}}$ - I(CII)${\rm_{HI}}$,
 %\begin{equation}
% I(CII)_{H_{2}} =  I(CII)_{total} - I(CII)_{HI}, 
% \end{equation}
we can now solve for N(C$^+$)${\rm_{H_2}}$, using equation (2), as a function of temperature and density (T$_{kin}$,n(H$_2$)). For illustrative purposes in  
Figure 2, 
we have chosen to consider N(H$_2$) a variable, and define the relative column densities of molecular and atomic gas through a parameter,   $\alpha$ = N(H$_2$)/N(HI).

In Figure 2
we plot I(CII) versus I(HI) for different values of $\alpha$ (dashed lines) ranging from 0 to 6 ($\alpha$ = 0 corresponds to a pure HI cloud), for a fixed temperature and density (T,n(HI)) = (100K,  300 cm$^{-3}$), and assume constant pressure and temperature, so n(H$_2$) = n(HI).  We assume that I(HI) yields the column density N(C$^+$) in the HI region, and N(H$_2$) is proportional to N(HI).  A modest layer of H$_2$ (N(H$_2$) = 1 to 2 $\times$ N(HI)) fits the distribution of many of the stronger [CII] clouds.  However, a much thicker layer of H$_2$ ($\alpha$ = 4 to 6) is required to explain the strongest [CII] emitting clouds in this sample.  

To characterize and compare the clouds and their dark gas content, we calculated the visual extinction A$\rm{_V}$ and ratio N(H$_2$)$/$N(HI) for each source, assuming a fixed (T, n) = (100K, 300 cm$^{-3}$); these are plotted in Figure 3 for each source.
We follow the procedure described above using N(HI) $\propto$ I(HI) to derive N(C$^+$)$\rm_{HI}$, then calculate the corresponding I(CII)${\rm _{HI}}$; subtract from I(CII)${\rm_{total}}$; finally, derive N(C$^+$)${\rm_{H_2}}$ using equation 2, and N(H$_2$) = N(C$^+$)${\rm_{H_2}}$$/$(2X(C$^+$)). then A$\rm{_V}$ = (N(HI) +2N(H$_2$))$\times$5.35$\times$10$^{-22}$.  The extinction ranges from 0.15 to 1.3 mag. and about 75\%  of the sources have N(H$_2$)$\ge$ 0.5N(HI), and about 30\% have N(H$_2$)$\ge$2N(HI). If the density and$/$or temperature in the H$_2$ layer is lower than assumed here, N(H$_2$) will be higher; and for higher (n,T) it will be lower. Here we assumed typical diffuse cloud conditions.  However, [CII] preferentially identifies warm gas, so some sources may be hotter and denser, perhaps in proximity to UV sources or sampling shocked gas, and have smaller N(H$_2$) than we calculate. Although it may the the case in a few individual clouds, statistically our sample of diffuse clouds observed in [CII] emission provides evidence for warm molecular ``dark gas" in diffuse regions.
%%%%%%%%%%%%%%%%%%%%55
\begin{figure}[t]
\includegraphics[width=0.45\textwidth,angle=0]{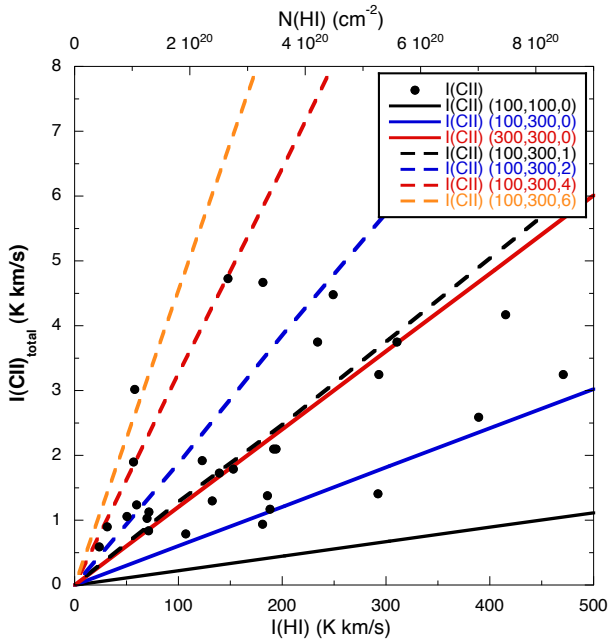}
\caption{Observed [CII] intensities (filled circles) versus I(HI) (lower axis); the corresponding N(HI) is given on the upper axis for optically thin lines.  Model calculations of I(CII) versus I(HI) as a function of (T$_{kin}$,n(HI),$\alpha$), where $\alpha$ is the ratio of N(H$_2$) to N(HI) in the cloud (see text). The labels (T,n,$\alpha$) for each curve are given in the box. The solid lines are for clouds in which all hydrogen is in HI; the bottom line represents typical diffuse HI conditions, while the top solid line is an upper limit for atomic clouds.  The dashed lines include H$_2$ gas in addition to HI, (see text); with values of $\alpha$=N(H$_2$)$/$N(HI) from 1 to 6.  Only about 50\% of the sources fit within the limits defined by the purely HI models, the remaining sources require significant N(H$_2$) to explain the large [CII] emission.  Most can be explained by N(H$_2$) equal to (1 - 4)$\times$N(HI). }
%\label{fig:15088fg2}
\end{figure} 

\section{Discussion}
\label{sec:discussion}

The suggestion of the presence of ``dark gas" in the interstellar medium is not new, Bloemen et al.\,(1986) used COS-B, Strong et al.\,(1996) EGRET, and Abdo et al.\,(2010) FERMI-LAT $\gamma$-ray data to show that the Galaxy has more gas mass than indicated by HI and CO surveys alone. Joncas et al.\,(1992) compared HI and 100 $\mu$m dust emission in infrared cirrus and from the excess infrared emission suggested there might be molecular hydrogen or cold atomic gas (T $<$ 30K) - see also Reach et al.\,(1994), who include CO. Grenier et al.\,(2005), who applied the term ``dark gas" to this component, extended this approach to include $\gamma$-ray observations and concluded that there were ``dark gas" layers (H$_2$ without CO) surrounding all the nearby CO clouds.  In addition, models of diffuse clouds and cloud envelopes (PDRs) predict significant amounts of H$_2$ gas with no CO, but containing C$^+$ and some neutral carbon, C$^o$. 

\begin{figure}[t]
\includegraphics[width=0.45\textwidth,angle=0]{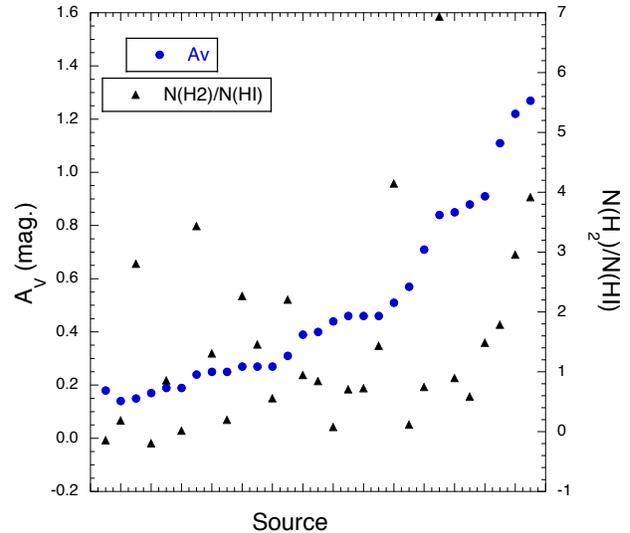}
\caption{Distribution of A$_{\rm V}$ and N(H$_2$)$/$N(HI) for all sources, based on a model calculation, assuming n(H$_2$) =300 cm$^{-3}$ and T$_{kin}$ = 100K. The sources are displayed with increasing A$\rm{_V}$.  For 2 sources, N(H$_2$)$/$N(HI) $<$ 0; a valid solution requires lower (n,T).}
%\label{fig:15088fg3}
\end{figure} 

Here we detected the warm ``dark gas" with [CII], and showed it is a tracer of this important ISM component.  Velusamy et al.\, (2010) use the $\bf{GOT\,C+}$ data base to analyze [CII] associated with $^{12}$CO clouds and find evidence for warm H$_2$ layers sandwiched between the HI and CO regions.  Thus, warm H$_2$ gas is an important component in interstellar clouds, in agreement with cloud-chemical models.  We estimate the [CII] emission from all diffuse components (this paper and transition clouds Velusamy et al.\, 2010 and find their sum to be about equal to that of the dense PDRs Pineda et al.\, 2010) along these 16 LOS.

\section{Conclusions}
\label{sec:conclusions}
We have detected, for the first time, warm ``dark gas" using the C$^+$ $^2$P$_{3/2}$--$^2$P$_{1/2}$ fine structure emission line at 1.9 THz (158 $\mu$m).  In just 16 lines-of-sight we detected 146 cloud components in [CII]  emission. Of these, 29 are diffuse clouds, and more than 75\% likely contain warm H$_2$. We analyzed the H$_2$ content and found that it ranges from a fraction of the HI cloud component, to being the dominant component, with N(H$_2$) = (1 to 4)$\times$N(HI). If T$_{kin}$ and n(H$_2$) are less than assumed in Figure 3, N(H$_2$) will be a larger fraction of the hydrogen in the cloud, and conversely for higher (n,T). Heterodyne observations of [CII] are an excellent probe for studying the warm ``dark gas"; they have the advantage of locating the gas in the Galaxy, and providing excitation and dynamical information about the clouds.  High spectral resolution observations will help us understand the relative contribution of [CII] from the diffuse ISM compared to dense PDRs associated with star forming regions. However, [CII] emission requires  T$_{kin}$ $\ge$ 30K,  and will miss cold HI and H$_2$ clouds. 
In conclusion,  the warm ``dark gas", is no longer ``dark", indeed it shines quite brightly in C$^+$.  
The full implementation of $\bf{GOT}$ $\bf{C+}$ will detect thousands of such clouds, and yield information for understanding how their properties depend on ISM conditions.

%\section{References}
%\label{sec:references}

\begin{acknowledgements}

This work was performed by the Jet Propulsion Laboratory, California
Institute of Technology, under contract with the National Aeronautics
and Space Administration. We thank the staffs of the ESA and NASA Herschel Science Centers for their help. The Mopra Telescope is managed by the Australia
Telescope, and funded by the Commonwealth of Australia for
operation as a National Facility by the CSIRO. We thank the referee and editor for comments.

\end{acknowledgements} 

%\bibliography{15088_references}
%\bibliographystyle{aa}

%\bibliography{/root/latex/papers}
%\bibliographystyle{/root/astronat/apj/apj.bst}

\clearpage

\end{document}